\documentclass[11pt,a4paper]{article}
\usepackage{jcapmod}
\usepackage{graphicx}
\usepackage{amsfonts}
\usepackage{subfigure}
\usepackage{hyperref}
\usepackage{color}        
\usepackage{amsmath}
\usepackage{empheq}
\usepackage{amssymb}
\usepackage{tikz}
\usepackage{tabularx}
\usepackage{booktabs}

\newcommand{\boxalign}[2][0.97\textwidth]{
 \par\noindent\tikzstyle{mybox} = [draw=black,inner sep=6pt]
 \begin{center}\begin{tikzpicture}
  \node [mybox] (box){%
   \begin{minipage}{#1}{\vspace{-5mm}#2}\end{minipage}
  };
 \end{tikzpicture}\end{center}
}

\def\d{{\rm d}}
\def\be{\begin{equation}}
\def\ee{\end{equation}}
\def\bea{\begin{eqnarray}}
\def\eea{\end{eqnarray}}
\def\ba{\begin{align}}

\def\bi{\begin{itemize}}
\def\ei{\end{itemize}}
\begin{document}

\vspace{2cm}
\begin{center}
{\fontsize{16}{28}\selectfont  \bf A Consistency Relation for the CMB B-mode Polarization in the Squeezed Limit}
\end{center}

\vspace{0.2cm}

\begin{center}
{\fontsize{13}{30} A. Kehagias$^{a,b}$, A. Moradinezhad-Dizgah$^{a}$, J. Nore\~na$^{a}$,  H. Perrier$^{a}$ and A. Riotto$^{a}$}
\end{center}

\begin{center}
\vskip 8pt
\textsl{$^{a}$ University of Geneva, Department of Theoretical Physics and Center for Astroparticle Physics (CAP), 24 quai E. Ansermet, CH-1211 Geneva 4, Switzerland }
\vskip 8pt
\textsl{$^{b}$ Physics Division, National Technical University of Athens,15780 Zografou Campus, Athens, Greece }
\end{center}

\vspace{1.24cm}
\hrule \vspace{0.3cm}
{ \noindent \textbf{Abstract} \\[0.2cm] 
A large-scale temperature perturbation has a non-zero correlation with the power spectrum of B-modes of cosmological origin on short scales while the corresponding correlation is expected to be zero for B-modes sourced by galactic foregrounds. We thus compute the squeezed limit of a three-point function in which one correlates the temperature fluctuations at large scales with two polarization modes at small scales. In the particular case of the B-mode polarization we obtain a relation that connects the squeezed limit of the $TBB$ three-point function with the cosmological B-mode power spectrum, which can be used as a consistency relation. This could in principle help to distinguish a primordial signal from that induced by inter-stellar dust. 
\noindent 
}  
\vspace{0.6cm}
 \hrule

\vspace{0.6cm}

\section{Introduction}

The recent detection of B-mode polarization pattern in the Cosmic Microwave Background (CMB) on degree-angular scales by the BICEP2 collaboration \cite{Ade:2014xna}, has generated a great deal of excitement in the field. The range of scales and the shape of the spectrum suggest that it has primordial origin and an outburst of work has followed up focusing on the implication of the reported signal for early universe physics. If this signal is indeed primordial it will be a strong argument in favor of inflation. However, it has been shown that this observation is also consistent with the polarized radiation emitted by the poorly-understood interstellar dust \cite{Flauger:2014qra, Mortonson:2014bja}. Nevertheless, these results open the possibility that a detailed observational study of B-mode polarization is not too far from our technological reach.  One may wonder whether measuring the three-point function in future experiments can help extract information contained in the B-mode signal. In particular, a large-scale temperature perturbation has a non-zero correlation with the power spectrum of B-modes of cosmological origin on short scales while the corresponding correlation is expected to be zero for B-modes sourced by galactic foregrounds.

Over the past decade the study of correlation functions of cosmological perturbations beyond the two-point function has received considerable attention. This interest was spurred in part by the fact that one can write consistency relations for primordial perturbations that relate $(n-1)$-point functions to $n$-point functions in the limit in which one of the scales is much larger than the others (the so-called squeezed limit) \cite{Maldacena:2002vr, Acquaviva:2002ud, Creminelli:2004yq, Creminelli:2011rh, Creminelli:2012ed,  Kehagias:2012pd, Assassi:2012zq, Hinterbichler:2013dpa}. The derivation of the consistency relations is based on the fact that a very long-wavelength mode of the gravitational potential should have no effect on quantities measured at much smaller scales and its effect is equivalent to a coordinate transformation. The consistency relations for primordial perturbations are useful for instance to distinguish whether the primordial perturbations are sourced by a single or multiple fields.  A similar argument can be used also to compute the squeezed limit of correlation functions involving actual observables, such as the CMB temperature \cite{Creminelli:2011sq, Bartolo:2011wb, Lewis:2012tc}  or the dark matter density \cite{Peloso:2013zw, Kehagias:2013yd, Creminelli:2013mca, Kehagias:2013xga, Kehagias:2013rpa, Valageas:2013cma, Creminelli:2013poa, Valageas:2013zda, Creminelli:2013nua, Nishimichi:2014jna, Horn:2014rta}.

In this note, we compute the squeezed limit of a three-point function in which one correlates the temperature fluctuations at large scales with two (E-mode or B-mode) polarization fluctuations at small scales using a method similar to the one used for deriving the consistency relations (in particular, we extend the approach of \cite{Creminelli:2011sq} to include polarization). To simplify the calculation, we do the computation in the flat-sky approximation and ignore late-time effects such as those induced by the late integrated Sachs-Wolfe (ISW) effect. The main contribution to the squeezed limit of such a correlation function comes from the lensing induced by the long-wavelength mode, and we estimate this signal to be observable for a futuristic experiment. This is similar to what happens for the CMB temperature bispectrum \cite{Boubekeur:2009uk, Lewis:2012tc}. An interesting feature of this result is that for the case of B-mode polarization we obtain a relation that connects the squeezed limit of the TBB three-point function with the primordial tensor power spectrum, which could in principle be used as a consistency relation for the tensor perturbations. Since the B-mode pattern sourced by inter-stellar dust is not expected to correlate with the temperature signal sourced at the CMB, this could help in distinguishing between dust and a primordial signal. It turns out, however, that the observation of this relation is difficult even for a very futuristic experiment. 

It is noteworthy that the full calculation of the temperature bispectrum in the absence of primordial non-Gaussianity can now be carried out numerically (see e.g. \cite{Huang:2013qua}). While this note was being prepared, an analytical and numerical study of the correlation function involving E-mode polarization appeared \cite{Pettinari:2014iha} and our results seem to agree with theirs.

We begin in Section \ref{sec:power} by computing the effect of a long-wavelength metric perturbation on the polarization power spectrum; we do this by closely following the approach of \cite{Creminelli:2011sq}. This two-point function can then be correlated with a long-wavelength temperature fluctuation to compute the squeezed limit of the temperature-polarization-polarization (TBB or TEE) bispectrum, which we perform in Section \ref{sec:bisp}. Finally, in Section \ref{sec:obs} we analyze whether such a bispectrum can be observable in an experiment similar to COrE, PRISM or an ideal noiseless experiment that probes very small scales. We draw conclusions in Section \ref{sec:conc}.

\section{The effect of the long scalar mode on the polarization power spectrum}
\label{sec:power}
In this section, we start by reviewing the effect of a long-wavelength background mode on the CMB temperature anisotropies as described in \cite{Creminelli:2011sq} (see also \cite{Bartolo:2011wb}) and then extend this treatment to the polarization field. 

The long-wavelength mode acts as a background for the short modes. We take the long mode to be constant at the scale of the short modes at recombination such that its effect is equivalent to a coordinate transformation ({\emph {i.e.} it is what Weinberg \cite{Weinberg:2003sw} calls an adiabatic mode) 
\bea \label{eq:co_trans}
\tilde \eta &=& \eta + \epsilon(\eta)\,, \nonumber \\
\tilde x^i &=& x^i(1-\lambda)\,,
\eea
where $\epsilon$ is an arbitrary function of time and $\lambda$ is an arbitrary constant. In the limit of instantaneous recombination and considering matter dominance, the observed CMB temperature anisotropies ignoring the ISW effect\footnote{The early ISW effect due to the evolution of Bardeen's potentials during the transition from radiation to matter domination has been taken into account since the standard calculation of the Sachs-Wolfe effect does not depend on this transition (see the discussion in Ref. \cite{Creminelli:2011sq}). So this approximation only ignores the late ISW effect due to the evolutions of the potentials in the presence of dark energy.} are given to first order by
\be\label{eq:temp_observ}
\Theta_{\rm obs}(\hat n) \equiv \left[\Theta + \Phi + \hat n\cdot \vec v \right](\eta_{\rm rec},\vec x_{\rm rec})\,,
\ee
In this expression, $\Theta$ is the intrinsic temperature anisotropy defined as 
\be
T(\eta, x^i) = \bar T(\eta)(1+\Theta(\eta,x^i))\,,
\ee
$\Phi$ is the gravitational potential and $\vec v$ is the velocity of the photon-baryon fluid. 
One can compute the effect of a long mode on the short scale temperature anisotropy by applying the coordinate transformation given in Eq. (\ref{eq:co_trans}) to each term in Eq. (\ref{eq:temp_observ}).
During matter-domination ($a \propto \eta^2$), we have $\epsilon =  \eta \ \Phi_L/3$ and $\lambda =5 \Phi_L/3$. 
Doing the coordinate transformation explicitly therefore gives
\be \label{eq:zeroth_order}
\Theta_{\rm obs} (\hat n)= \Theta_{{\rm obs}, S} (\hat n) + \Theta_{{\rm obs},L} (\hat n) + \Theta_{{\rm obs},L} (\hat n)\left(\frac{\partial }{\partial \ {\rm ln} \ \eta_{\rm rec}} + 1 - 5\ \hat n\cdot \nabla_{\hat n} \right)\Theta_{{\rm obs},S}(\hat n)\,.
\ee
Note that this expression is valid when the long mode is taken to be larger than the sound horizon at recombination but the short modes can be taken to be at arbitrarily small scales, as it relies only on the fact that a constant gravitational potential has no physical effect and is therefore equivalent to a coordinate transformation. Also note that this expression is indeed compatible with the explicit second order calculation performed e.g. in references \cite{Bartolo:2006cu, Bartolo:2006fj}. The time derivative in equation \eqref{eq:zeroth_order} will be suppressed by $\eta_{\rm rec}/\eta_{\rm obs}$ and will thus be subdominant with respect to the rescaling.

The presence of the long-wavelength mode will also change the relation between the direction of the observation $\hat n$ and the physical position at recombination $\vec x_{\rm rec}$. In the absence of the long-wavelength mode they are related through the zeroth order geodesic equation, $\vec x_{\rm rec} = \hat n(\eta_{\rm obs} - \eta_{\rm rec})$. In writing  Eq. (\ref{eq:zeroth_order}) this relation is used to rewrite $\vec x_{\rm rec} \cdot\nabla_{\vec x_{\rm rec}} = \hat n\cdot \nabla_{\hat n}$. In the presence of the long-wavelength mode the relation between $\vec x$ and $\hat n$ is modified as
\begin{multline}
\label{eq:pos_ch}
\vec x = \hat n \left[\left(1-\frac{1}{3} \Phi_L(x_{\rm obs})\right)\eta_{\rm obs} - \left(1-\frac{1}{3} \Phi_L (x_{\rm rec})\right)\eta_{\rm rec}\right] \\ + 2 \hat n \int_{\eta_{\rm rec}}^{\eta_{\rm obs}} \Phi_L(\vec x) d \eta - 2 \int_{\eta_{\rm rec}}^{\eta_{\rm obs}} \d \eta (\eta-\eta_{\rm rec})\vec \nabla_\perp \Phi_L(\vec x)\,.
\end{multline}
This relation is obtained by solving the photon geodesic equation at first order. As described in \cite{Creminelli:2011sq}, only the last term of the above expression, which is the lensing term, will contribute to the bispectrum. The first term in the square brackets does not contribute since it only depends on the gravitational potential at observation. The second term in the square bracket changes the distance to the last scattering surface and is suppressed by a factor of $\eta_{\rm rec}/\eta_{\rm obs}$. The second term is also suppressed since the integral of $\Phi$ along the line of sight, the so called Shapiro effect, tends to average out along the line of sight for high enough multipoles.

Therefore the effect of the long-wavelength mode on the short modes reduces to the lensing induced by the long mode (the last term in Eq. (\ref{eq:pos_ch})) and a stretching perpendicular to the line of sight (the last term in Eq. (\ref{eq:zeroth_order})). From now on for simplicity, we drop the subscript {\it observed} in the temperature anisotropy. The short-wavelength mode, $\Theta_S$, in the presence of long-wavelength mode, $\Theta_L$, is modified as
\be\label{eq:short_temp}
\Theta(\hat n) = \Theta_S \left(\hat n + \vec \alpha(\hat n)\right) + \Theta_L(\hat n)\left( 1 - 5\ \hat n\cdot \nabla_{\hat n} \right)\Theta_S(\hat n),
\ee
where $\vec \alpha$ is the deviation due to the lensing term and we dropped the subscript $S$ for the short modes. 

We can similarly find the effect of a long-wavelength temperature mode on the short-scale polarization mode. In the presence of the long mode, the short-wavelength polarization field $\Theta_P$ will be 
\be
\Theta_P(\hat n) = \Theta_{P,S}(\hat n + \vec \alpha(\hat n)) - 5 \Theta_L \hat n\cdot \nabla_{\hat n} \Theta_{P,S}(\hat n), 
\ee
Notice that the constant term in Eq. (\ref{eq:short_temp}), which corresponds to change in the average temperature of CMB, does not affect the polarization field. This is because it corresponds to the monopole of the temperature field and the polarization is only sensitive to the quadruple moment. Using the flat-sky approximation $\hat n \sim (1,\vec m)$ we get the power spectrum of the two short polarization modes in the presence of a long-wavelength temperature mode
\bea\label{eq:pol_power}
\langle  \Theta_P(\vec m_1 )  \Theta_P (\vec m_2)  \rangle_L &=& \langle \Theta_P(\vec m_1) \Theta_P(\vec m_2) \rangle + (\alpha^i(\vec m_1) - \alpha^i(\vec m_2)) \nabla_i\langle \Theta_P(\vec m_1) \Theta_P(\vec m_2)\rangle \nonumber \\
&-& 5 \Theta_L \vec m\cdot\nabla_{\vec m} \langle \Theta_P(\vec m_1)\Theta_P(\vec m_2)\rangle.
\eea
where we have used the fact that the power spectrum depends only on the distance $\vec m = \vec m_1 - \vec m_2$, which is small due to the fact that we are in the flat-sky or distant observer approximation. From now on we will ignore the first term which does not contribute to the bispectrum. 

Next, by decomposing the polarization field into $E$ and $B$ modes we calculate how the presence of the long-wavelength temperature mode modifies the polarization power spectrum. This will then lead us to the calculation of the TBB and TEE bispectra in the next section. The polarization matrix for linearly polarized radiation is a real spin-2 object parametrized in terms of Stokes parameters $Q$ and $U$ 
\begin{equation}
P = \left(\begin{array}{cc}
Q & U \\
U & -Q 
\end{array}\right).
\end{equation}
Under a counterclockwise rotation through an angle $\phi$, the stokes parameters transform as
\begin{align}
Q' &= Q\cos 2\phi  + U\sin 2\phi,  \nonumber \\
U' &= U\cos 2\phi - Q\sin 2\phi,
\end{align}
or in a more compact form
\be
(Q \pm i U )' = e^{\mp2i\phi}(Q \pm i U),
\ee
which indicates that the polarization field $\Theta_P= Q \pm iU$ is a spin-$2$ quantity. Note that $Q$ is parity even while $U$ is parity odd. Putting together quantities of the same parity, we can construct spin-0 quantities $E$ and $B$ by applying the spin raising and lowering operators on $Q$ and $U$. In the flat-sky approximation, where one neglects the curvature of the sphere and consider it as a plane normal to ${\textbf e}_z$ the spin raising and lowering operators reduce to 
\be\label{eq:operators}
L_\pm = L_x \pm i L_y = \partial_x \pm i\partial_y.
\ee
 Therefore the spin-0 quantities $\tilde E$ and $\tilde B$ can be defined as \cite{Zaldarriaga:1996xe}
\begin{align}
\tilde{E} &= -\frac{1}{2}\left(L_{+}^2 (Q - i U) + L_{-}^2 (Q + i U)\right)\,, \label{eq:etilda} \\
\tilde{B} &=\frac{1}{2i} \left(L_{+}^2 (Q - i U) - L_{-}^2 (Q + i U)\right)\,, \label{eq:btilda}
\end{align}
and under a Fourier transform
\begin{align}
a_{\tilde{E}}(\vec{l}) &= \int \mathrm{d}\vec{m}\, l^2 (Q\cos2\phi_l + U\sin 2\phi_l)e^{-i\vec{l}\cdot\vec{m}}, \\
a_{\tilde{B}}(\vec{l}) &= \int \mathrm{d}\vec{m}\, l^2 (-Q\sin2\phi_l + U\cos 2\phi_l)e^{-i\vec{l}\cdot\vec{m}}\,.
\end{align}
We will write our final results in terms of a rescaled coefficients $a_{(E,B)} = a_{(\tilde E,\tilde B)}/l^2$.

Next we use Eqs. (\ref{eq:operators}) and (\ref{eq:btilda}) to compute the effect of a rescaling on the $B$ modes. First we consider the last term in Eq. (\ref{eq:pol_power}), putting aside the lensing term for now. Note that
\begin{equation}
L_\pm^2 \vec{m}\cdot\nabla_{\vec{m}} = (2 + \vec{m}\cdot\nabla_{\vec{m}}) L_\pm^2\,,
\end{equation}
where we used the fact that $L_\pm^2$ is given simply by a combination of second derivatives. Therefore the contribution from the last term to the polarization is given by
\begin{align}
\tilde{B}(\vec{m}) \mapsto &\tilde{B}(\vec{m})-  5 \Theta_L\big[-i L_{+}^2 \vec{m}\cdot\nabla_{\vec{m}} (Q - i U) + i L_{-}^2 \vec{m}\cdot\nabla_{\vec{m}} (Q + i U)\big]  \nonumber \\
&=\tilde{B}(\vec{m})-5\Theta_L(2 + \vec{m}\cdot\nabla_{\vec{m}})\tilde{B}(\vec{m})\,
\end{align}
and analogously for the E modes
\begin{equation}
\tilde{E}(\vec{m}) \mapsto \tilde{E}(\vec{m})-5\Theta_L(2 + \vec{m}\cdot\nabla_{\vec{m}})\tilde{E}(\vec{m})\,.
\end{equation}
Hence its contribution to the polarization power spectrum is given by
\begin{align}\label{eq:Bpower_rescale}
\langle a_{\tilde{X}}(\vec{l}_1) a_{\tilde{X}}(\vec{l}_2)\rangle_L \  {\supset} &- \int \mathrm{d}\vec{m}_1\mathrm{d}\vec{m}_2\,e^{-i(\vec{l}_1\cdot\vec{m}_1 + \vec{l}_2\cdot\vec{m}_2)} \nonumber\\
&\times 5\Theta_L(\vec{M})(4 + \sum_i\vec{m}_i\cdot\nabla_{\vec{m}_i})\langle \tilde{X}(\vec{m}_1)\tilde{X}(\vec{m}_2)\rangle, 
\end{align}
where $\vec{M} \equiv (\vec{m}_1 + \vec{m}_2)/2$ and $X$ stands for either $E$ or $B$. 

Next, we consider the contribution from the lensing piece, which is given by
\begin{align}
 \Theta_P(\vec m + \vec \alpha(\vec m)) &= (Q \pm i U)(\vec{m} + \vec{\alpha}(\vec{m})) \nonumber \\
&\simeq (Q \pm iU)(\vec{m}) + \alpha^i(\vec{M})\nabla^i_m (Q \pm i U)(\vec{m}) \nonumber \\
&\phantom{=} + (m^j - M^j)\nabla^j_M \alpha^i(\vec{M})\nabla^i_m (Q \pm i U)(\vec{m})\,,
\label{eq:qpu}
\end{align}
where in the last line we have assumed that $\alpha$ varies slowly since it is given by the long-wavelength mode and we have Taylor expanded it around $\vec{M}$. Since the polarization modes are combinations of $L_\pm^2$ acting on these objects, let us act on the above equations with the raising and lowering operators. Also let us focus only on the piece in the last line proportional to $\vec{m}$, since it is the only non-trivial one
\begin{align}
L_\pm^2 m^j \nabla^j_M \alpha^i(\vec{M})\nabla^i_m (Q \pm i U)(\vec{m}) &=  m^j \nabla^j_M \alpha^i(\vec{M})\nabla^i_m L_\pm^2(Q \pm i U)(\vec{m}) \nonumber \\
&+ (L_\pm^2 \nabla_M^2 + L_\pm^{(M)\;2}\nabla_m^2)\psi(M)(Q \pm i U)(\vec{m})\,,
\label{eq:nontrivial}
\end{align}
where we write the derivatives with respect to $M$ in terms of raising and lowering operators written in terms of the capital coordinates $L_\pm^{(M)} = \pm(\partial_{M_1} \pm i \partial_{M_2})$ with  $\vec{M} = (M_1, M_2)$. By putting together Eqs. (\ref{eq:btilda}, \ref{eq:qpu}, \ref{eq:nontrivial}) we get
\begin{align}
\tilde{B}(\vec{m}) \mapsto \bigg(1 &+  \nabla^i_M\psi(\vec{M})\nabla^i_m - M^j\nabla^j_M\nabla^i_M\psi(\vec{M})\nabla_m^i  \nonumber \\
 &+ m^j\nabla^j_M\nabla^i_M\psi(\vec{M})\nabla_m^i  + \nabla^2_M\psi(\vec{M})\bigg) \tilde{B}(\vec{m}) \nonumber\\ 
 &-\frac{i}{2}L_+^{(M)\;2}\psi(\vec{M})\nabla_m^2(Q - iU)(\vec{m}) + \frac{i}{2} L_-^{(M)\;2}\psi(\vec{M})\nabla_m^2(Q + i U)(\vec{m})\,,
\end{align}
where only the second and third line will contribute to the final answer. Using the definitions of Eqs. (\ref{eq:etilda}) and (\ref{eq:btilda}), it's easy to rewrite the third line as
\begin{equation}
\frac{i}{2}L_+^{(M)\;2}\psi(\vec{M})\frac{L_-^2}{\nabla_m^2}(\tilde{E} - i\tilde{B})(\vec{m}) - \frac{i}{2} L_-^{(M)\;2}\psi(\vec{M})\frac{L_+^2}{\nabla_m^2}(\tilde{E} + i \tilde{B})(\vec{m})\,.
\end{equation}
Analogously for the E-modes,
\begin{align}
\tilde{E}(\vec{m}) \mapsto \bigg(1 &+  \nabla^i_M\psi(\vec{M})\nabla^i_m - M^j\nabla^j_M\nabla^i_M\psi(\vec{M})\nabla_m^i  \nonumber \\
 &+ m^j\nabla^j_M\nabla^i_M\psi(\vec{M})\nabla_m^i  + \nabla^2_M\psi(\vec{M})\bigg) \tilde{E}(\vec{m}) \nonumber\\ 
 &+\frac{1}{2}L_+^{(M)\;2}\psi(\vec{M})\nabla_m^2(\tilde{E} - i\tilde{B})(\vec{m}) + \frac{1}{2} L_-^{(M)\;2}\psi(\vec{M})\nabla_m^2(\tilde{E} + i \tilde{B})(\vec{m})\,.
\end{align}
Therefore the contribution of the lensing term to the polarization power spectrum in the presence of long-wavelength temperature mode is given by 
\begin{multline}
\langle a_{\tilde{X}}(\vec{l}_1) a_{\tilde{X}}(\vec{l}_2)\rangle_L \ {\supset} \int \mathrm{d}\vec{m}_1\mathrm{d}\vec{m}_2\,e^{-i(\vec{l}_1\cdot\vec{m}_1 + \vec{l}_2\cdot\vec{m}_2)} \\
\times \bigg[2\nabla^2_{M} + \nabla_{M}^i\nabla_{M}^j m^i \nabla^j_m + L_+^{(M)\;2}L_-^2\nabla^{-2}_m + L_-^{(M)\;2}L_+^2\nabla^{-2}_m\bigg]\psi(\vec{M})\langle \tilde{X}(\vec{m}_1)\tilde{X}(\vec{m}_2)\rangle\,.
\label{eq:Bpower_lens}
\end{multline}

\section{The bispectrum in the squeezed limit}
\label{sec:bisp}
Having calculated the polarization power spectrum in the presence of long-wavelength temperature mode given by Eqs. (\ref{eq:Bpower_rescale},  \ref{eq:Bpower_lens}), we can finally calculate the TBB and TEE bispectra in the squeezed limit by correlating this power spectrum with the long-wavelength temperature mode.  Again let's first consider only the contribution from the rescaling part. Using the fact that the two-point function depends only on $\vec{m} \equiv \vec{m}_1 - \vec{m}_2$ we get 
\begin{align}
\langle a_{\tilde{X}}(\vec{l}_1) a_{\tilde{X}}(\vec{l}_2) a(\vec{l}_3)\rangle \overset{\vec{l}_3 \rightarrow 0}{\supset}  &-\int\mathrm{d}\vec{m}\mathrm{d}\vec{M}\mathrm{d}\vec{M}_L\,e^{-i((\vec{l}_1 + \vec{l}_2)\cdot\vec{M} + \vec{l}_3\cdot\vec{M}_L + (\vec{l}_1 - \vec{l}_2)\cdot\vec{m}/2)} \nonumber\\
&\phantom{=}\times 5\langle\Theta(\vec{M}_L)\Theta_L(\vec{M})\rangle(4+\vec{m}\cdot\nabla_{\vec{m}})\langle \tilde{X}(\vec{m}_1)\tilde{X}(\vec{m}_2)\rangle \nonumber \\
& = (2\pi)^2\delta(\vec{l}_1 + \vec{l}_2 + \vec{l}_3) 
5\big(-2 + \vec{l}_1\cdot\nabla_{l_1}\big)C^{\tilde{X}\tilde{X}}(l_1)C(l_3)\,.
\end{align}
Now we want to write the corresponding relation for $a_X$ (without the tilda), remembering that $a_X = a_{\tilde{X}}/l^2$, that is
\begin{align}
\langle a_X(\vec{l}_1) a_X(\vec{l}_2) a(\vec{l}_3)\rangle \overset{\vec{l}_3 \rightarrow 0}{\supset} &(2\pi)^2\delta(\vec{l}_1 + \vec{l}_2 + \vec{l}_3) 
\frac{5}{l_1^4}\big(-2 + \vec{l}_1\cdot\nabla_{l_1}\big)(l_1^4 C^{XX}(l_1))C(l_3) \nonumber \\
&= (2\pi)^2\delta(\vec{l}_1 + \vec{l}_2 + \vec{l}_3) 
5C^{XX}(l_1) C(l_3) \frac{\mathrm{d}\ln \left(l_1^2 C^{XX}(l_1)\right)}{\mathrm{d}\ln l_1}.
\end{align}

Next we consider the contribution of lensing term to the bispectrum by correlating Eq.(\ref{eq:Bpower_lens}) with the long temperature mode. Let's consider each term in the square parenthesis separately:
the first term in the square parenthesis of Eq. (\ref{eq:Bpower_lens}) is just trivially computed to be
\be
\langle a_{\tilde{X}}(\vec{l}_1) a_{\tilde{X}}(\vec{l}_2) a(\vec{l}_3)\rangle \overset{\vec{l}_3 \rightarrow 0}{\supset} -(2\pi)^2\delta(\vec{l}_1 + \vec{l}_2 + \vec{l}_3) 2 l_3^2 C^{\tilde{X}\tilde{X}}(l_1)C^{T\psi}(l_3)\,.
\ee
The second term, which has a similar structure to the lensing of the temperature power spectrum, is
\begin{align}
\langle a_{\tilde{X}}(\vec{l}_1) a_{\tilde{X}}(\vec{l}_2) a(\vec{l}_3)\rangle \overset{\vec{l}_3 \rightarrow 0}{\supset} &\int\mathrm{d}\vec{m}\mathrm{d}\vec{s}\mathrm{d}\vec{S}\,e^{-i((\vec{l}_1 + \vec{l}_2 + \vec{l}_3)\cdot\vec{S} - (\vec{l}_1 + \vec{l}_2 - \vec{l}_3)\cdot\vec{s}/2 + (\vec{l}_1 - \vec{l}_2)\cdot\vec{m}/2)} \nonumber \\ 
&\phantom{=}\times  m^i\nabla^i_{\vec{s}}\nabla^j_{\vec{s}}\langle\Theta(\vec{M}_L)\psi(\vec{M})\rangle\nabla_{\vec{m}}^j\langle\tilde{X}(\vec{m}_1)\tilde{X}(\vec{m}_2)\rangle \nonumber \\
& = -(2\pi)^2\delta(\vec{l}_1 + \vec{l}_2 + \vec{l}_3)  l_3^2 C^{\tilde{X}\tilde{X}}(l_1)C^{T\psi}(l_3) \nonumber \\
&\phantom{=} \times \Bigg[\cos(2\varphi) - \cos^2\varphi \frac{\mathrm{d}\ln\left(l_1^2 C^{\tilde{X}\tilde{X}}(l_1)\right)}{\mathrm{d}\ln l_1}\Bigg]\,,
\end{align}
where $\varphi$ is the angle between the vectors $\vec{l}_1$ and $\vec{l}_3$. Finally, the third and fourth terms in the square parenthesis of Eq. (\ref{eq:Bpower_lens}) have a more complicated structure but their computation is straightforward
\begin{align}
\langle a_{\tilde{X}}(\vec{l}_1) a_{\tilde{X}}(\vec{l}_2) a(\vec{l}_3)\rangle \overset{\vec{l}_3 \rightarrow 0}{\supset} &-(2\pi)^2\delta(\vec{l}_1 + \vec{l}_2 + \vec{l}_3) C^{\tilde{X}\tilde{X}}(l_1)C^{T\psi}(l_3) \nonumber \\ 
&\times \frac{2}{l_1^2}\Big[\big((l_3^x)^2-(l_3^y)^2\big)\big((l_1^x)^2 - (l_1^y)^2\big) + 4l_1^x l_1^y l_3^x l^y_3\Big] \nonumber \\
&= -(2\pi)^2\delta(\vec{l}_1 + \vec{l}_2 + \vec{l}_3) 2 l_3^2 \cos(2\varphi) C^{\tilde{X}\tilde{X}}(l_1)C^{T\psi}(l_3)\,.
\end{align}
Putting everything together we obtain the lensing contribution to the bispectrum
\begin{align}
\langle a_{\tilde{X}}(\vec{l}_1) a_{\tilde{X}}(\vec{l}_2) a(\vec{l}_3)\rangle \ {\supset}&-(2\pi)^2\delta(\vec{l}_1 + \vec{l}_2 + \vec{l}_3)  l_3^2 C^{\tilde{X}\tilde{X}}(l_1)C^{T\psi}(l_3) \nonumber \\ 
&\times \left[2 + 3\cos(2\varphi) - \cos^2\varphi \frac{\mathrm{d}\ln\left(l_1^2 C^{\tilde{X}\tilde{X}}(l_1)\right)}{\mathrm{d}\ln l_1}\right]\,.
\end{align}
Finally, after changing from $\tilde{X}$ to $X$, we get
\begin{align}
\langle a_X(\vec{l}_1) a_X(\vec{l}_2) a(\vec{l}_3)\rangle \overset{l_3 \rightarrow 0}{\supset}&-(2\pi)^2\delta(\vec{l}_1 + \vec{l}_2 + \vec{l}_3)  l_3^2 C^{XX}(l_1)C^{T\psi}(l_3) \nonumber \\
 &\times \left[\cos(2\varphi) - \cos^2\varphi \frac{\mathrm{d}\ln\left(l_1^2 C^{XX}(l_1)\right)}{\mathrm{d}\ln l_1}\right].
\end{align}
Adding the rescaling and lensing contributions to the bispectrum given in Eqs. (\ref{eq:Bpower_rescale}, \ref{eq:Bpower_lens}),  we obtain the TBB and TEE bispectrum in the squeezed limit where the temperature mode has a much longer wavelength than the two polarization modes.
\boxalign[0.9\textwidth]{
\begin{align}\label{eq:TXX}
\langle a_X(\vec{l}_1) a_X(\vec{l}_2) a(\vec{l}_3)\rangle &\overset{l_3 \rightarrow 0}{=}(2\pi)^2\delta(\vec{l}_1 + \vec{l}_2 + \vec{l}_3)  l_3^2 C^{XX}(l_1) \nonumber \\ &\phantom{=}\times \Bigg[C^{T\psi}(l_3)\Bigg(-\cos(2\varphi) + \cos^2\varphi \frac{\mathrm{d}\ln\left(l_1^2 C^{XX}(l_1)\right)}{\mathrm{d}\ln l_1}\Bigg) \nonumber \\  &\phantom{=\times\Bigg[}+ 5 C^{TT}(l_3) \frac{\mathrm{d}\ln \left(l_1^2 C^{XX}(l_1)\right)}{\mathrm{d}\ln l_1}\Bigg]\,,
\end{align}
}
where $X$ denotes either $E$ or $B$.

Note that in Eq. \eqref{eq:TXX}, the logarithmic derivative will be sensitive to the tilt of the primordial power spectra. In particular, when computing the TBB bispectrum, it will receive a contribution proportional to the tilt of the tensor power spectrum $n_T$. However, as we will see in the next section it is difficult, to say the least, to observe this contribution. 

\section{Signal-to-noise estimation}
\label{sec:obs}
Similar to the temperature bispectrum \cite{Hu:2000ee}, for the TBB and TEE bispectra given in Eq. \eqref{eq:TXX}, the signal-to-noise is given by \begin{equation}
\left(\frac{S}{N}\right)^2 = \frac{f_{\rm sky}}{\pi}\frac{1}{(2\pi)^2} \iint  {\rm d}^2 l_1\ {\rm d}^2 l_3 \ \frac{\left({B_{l_1l_2l_3}^{\rm TXX}}\right)^2}{{\rm Var}},
\end{equation}
where again X can be either E or B. Assuming $l_3<l_2<l_1 $, the variance is given by
\begin{equation}
{\rm Var} = \langle \tilde a_T^*(l_3) \tilde a_X^*(l_1) \tilde a_X^*(l_2) \tilde a_T(l_3) \tilde a_X(l_1) \tilde a_X(l_2) \rangle \approx \tilde C^{TT}(l_3) \tilde C^{XX} (l_1) \tilde C^{XX}(l_2).
\end{equation}
The spectra with a tilde are the theoretical power spectra plus the instrumental noise
\begin{equation}
\tilde C^{YY}_l = C^{YY}_l + N_l^{YY},
\end{equation}
where Y can be T, E or B mode.  The noise power spectrum for a multi-frequency experiment like Planck is given by  \cite{Bowden:2003ub}
\begin{equation}
N_l^{YY} = \left(\sum_c \frac{1}{N_{l,c} ^{YY}}\right)^{-1}.
\end{equation}
The noise in each channel is given by
\begin{equation}
N_{l,c}^{YY} = \theta_{\rm fwhm}^2 \sigma_Y^2 \  {\rm exp} \left[ l(l+1)\frac{\theta_{\rm fwhm}^2}{8{\rm ln}2}\right],
\end{equation}
where $\theta_{\rm fwhm}$ is the full width at half maximum of the Gaussian beam and $\sigma_Y$ is the root mean square of the instrumental noise. Non-diagonal noise terms are supposed to vanish since the noise contribution from different maps are uncorrelated.
In our signal-to-noise estimation we use the values of $\theta_{\rm fwhm}$ and $\sigma_Y$ for three frequency channels of the Planck 14-month mission 
\cite{Planck:2006aa}, seven frequency channels of the COrE 4-year mission \cite{Bouchet:2011ck}, five frequency channels of the PRISM 4-year mission \cite{Andre:2013afa} as given in Tables [2-4] of the Appendix and an ideal noiseless experiment with $l_{\rm max} = 3000$. For all four cases we take the sky fraction to be $f_{\rm sky} = 0.65$. For the fiducial model, we consider a 6 parameter cosmology with $\{A_s =2.215 \times 10^{-9}, \ \Omega_m = 0.1199, \ \Omega_b = 0.02205, \  \tau =0.0925, \ n_s = 0.962, \ r= 0.1\}$  and $n_T=-(r/8)(2-r/8-n_s)$ satisfying the single-field inflationary consistency relation. We evaluate the integrals in the squeezed limit, we choose the long-wavelength mode to be in the range $20\leq l_3\leq 300$ and the two short-wavelength modes to be equal and in the range $10 \ l_3 \leq l_1 \leq l_{\rm max}$. The power spectra are computed by the numerical code CLASS \cite{Lesgourgues:2011re}.

The signal-to-noise of the TBB and TEE bispectra in the squeezed limit for these three experiments and a zero noise experiment are given in table [\ref{tab:S2N}]. We quote two cases: considering only the rescaling part of Eq. \eqref{eq:TXX}, and using the full formula. The signal is dominated by the lensing induced by the long mode. In principle this can be subtracted from observations; for the TEE correlation the resulting signal would still be observable in the admittedly far-fetched noiseless experiment, while for the TBB correlation the signal-to-noise would barely be greater than one even for such a futuristic experiment. In particular, the contribution to TBB coming from the tilt of the primordial tensor power spectrum is inaccessible to a direct measurement. The E-modes signal-to-noise is larger as it is also sourced by temperature anisotropies. Comparing with \cite{Pettinari:2014iha}, we get compatible values taking into account that we have differences in our computations: on the one hand we only include triangles in the squeezed limit whereas they include all triangles which tends to lower our signal-to-noise although most of their signal comes form squeezed configurations. 

\setlength{\extrarowheight}{4pt}
\begin{table}
\centering
\begin{tabular}{c c | c c c c}
Modes & long-mode lensing & Planck & COrE & PRISM& Ideal \\
\hline
TBB & no & $6.7 \times 10^{-5}$ & $2.4 \times 10^{-2}$ & $5.0 \times 10^{-2}$ &$1.1$ \\
TBB & yes &$9.8 \times 10^{-4}$ & $0.40$ & $0.72$ &$8.7$ \\
TEE & no & $0.16$ & $1.7$ & $2.3$ &$3.1$ \\
TEE & yes & $1.8$ & $14$ & $18$ &$23$ \\
\end{tabular}
\caption{Bispectrum signal-to-noise.}
\label{tab:S2N}
\end{table}

\section{Conclusions}
\label{sec:conc}

We have computed the squeezed limit of the correlation function involving one temperature and two polarization fluctuations. This has been done by appealing to the fact that a constant gravitational potential (or metric fluctuation in the Poisson gauge) can have no physical effect on the local observables. It is worth noting that we do not expect interstellar dust to give any contribution to Eq. \eqref{eq:TXX}, since, being within our galaxy, it should not correlate with a long-wavelength fluctuation at the CMB.

Our results indicate that a direct observation of this squeezed limit for the B-mode polarization is possible only for very futuristic experiments, while its observation for E-mode polarization is more plausible. As pointed out in \cite{Pettinari:2014iha}, this effect has to be correctly taken into account if one is to use the E-mode bispectrum in order to constrain primordial non-Gaussianity. 

One could have hoped to use the B-mode bispectrum to learn something about its the nature of the primordial universe. An example of this is the dependence of the bispectrum on the tilt of the primordial tensor power spectrum, which would then be expected to be compatible with the tilt measured in the B-mode two-point function. However, most of the signal comes from the lensing induced by the long-wavelength mode, an effect that contains little information about the shape of the primordial bispectrum. Even if one were to subtract this lensing effect, the contribution of the lensing to the variance would still hamper observations. An adventurous alternative would be to measure the variation of the B-mode power spectrum among different small patches in the sky and correlate this with the average temperature of each patch, similar to what was proposed for the large scale structure in ref. \cite{Chiang:2014oga}. One could then attempt to subtract the lensing in each patch using lensing potential maps in order to reduce the lensing variance. We leave a detailed analysis of such a technique as future work if it proves interesting. 

\section*{Acknowledgements}

It is a pleasure to thank M. Kunz for helpful conversations and M. Tucci for interesting discussions on galactic foregrounds. H.P., J.N. and  A.R. are supported by the Swiss National Science Foundation (SNSF), project `The non-Gaussian Universe" (project number: 200021140236). The research of A. M. is supported by the Tomalla foundation for Gravity Research. The  research of A.K. was implemented under the ``Aristeia" Action of the 
``Operational Programme Education and Lifelong Learning''
and is co-funded by the European 
Social Fund (ESF) and National Resources.  This work is partially
supported by European Union's Seventh Framework Programme (FP7/2007-2013) under REA
grant agreement n. 329083. 

\appendix
\section{ Applying the coordinate transformations directly to BB power spectrum} 
In this appendix we derive the rescaling contribution to the B-mode power spectrum given in  Eq. (\ref{eq:Bpower_rescale}), by applying the rescaling directly to the B-mode power spectrum in Fourier space. We will confirm that that it is given by
\be\label{eq:resc}
C_l \longmapsto C_l\left( 1+ 5\, a(l_3) \frac{{\mathrm{d}} \ln l^2 C_l}{{\mathrm{d} }\ln l}\right)\,, 
\ee
where $a(l_3)$ is the multipole moment of the long mode temperature perturbation, which is related to the gravitational potential by $\Theta = (T-\bar{T})/\bar{T} = \Phi/3$.
During matter-domination, the effect of the long-wavelength mode corresponds to a rescaling of the wave vector $\vec k$ and the time coordinate $\eta$ given by
\begin{equation}\label{eq:rescale}
\vec{k} \longmapsto \vec{k}\,e^{5 \Phi_L /3}, \hspace{1cm} \eta \longmapsto \eta \,e^{ \Phi_L /3}.
\end{equation}
The BB power spectrum is given by \cite{Gorbunov:2011zzc}
\be\label{eq:CBB}
C_l^{BB} = \frac{4 \pi}{25} \Delta \eta_{\rm rec}^2 \int_0^{\infty} \d \ln k\,\, P_T(k) \mathfrak{h}^{'2}(k,\eta_{\rm rec})\left[ \frac{l+2}{2l+1}j_{l-1}(k \eta_{\rm obs})-\frac{l-1}{2l+1}j_{j+1}(k \eta_{\rm obs})\right]^2
\ee
with primordial tensor power spectrum given by
\begin{equation}
P_T(k)=A_T \left(\frac{k}{k_*}\right)^{n_T},
\end{equation}
and transfer function is
\begin{equation}
\mathfrak{h}'(k,\eta_{\rm rec}) = - 3 \frac{j_2(k \eta_{\rm rec})}{\eta_{\rm rec}}.
\end{equation}
The term in the square brackets can be rewritten in terms of the Bessel function and its derivative. At $l \gg 1$ the asymptotic behavior of spherical Bessel function is given by a cosin function. Therefore since the phase of the Bessel function and its derivative differs by $\pi/2$ the cross term that is proportional to $j_l\cdot j'_l$ can be neglected. Averaging the oscillating function of $j_l$ and $j'_l$ one gets
\begin{equation}
C_l^{BB} \simeq \frac{36 \pi}{25} \Delta \eta_{\rm rec}^2 \int_{(l+1/2)/ \eta_{\rm obs}}^{\infty} d \ln k \,\, P_T(k) \frac{j_2^2(k \eta_{\rm rec})}{\eta_{\rm rec}^2} \left[ \frac{\sqrt{(k\eta_{\rm obs})^2-(l+1/2)^2}}{2(k\eta_{\rm obs})^3}\right].
\end{equation}
There are some subtleties in applying the above rescaling to the B-mode power spectrum as the rescaling should be applied on the power spectrum at recombination (primordial power spectrum times the transfer function) but not  on the geometrical projection effect, in this case given the square brackets in Eq. (\ref{eq:CBB}). One can also ignore any rescaling of the time at the observer because it is not observable. Finally, and most importantly, the addition of a long mode to the gravitational potential locally increases the average temperature as
\begin{equation}
\bar{T} \rightarrow \bar{T}e^{\Phi_L/3}.
\end{equation}
Recombination happens however at a fixed physical temperature. In the patch at recombination where the long mode is present, recombination is therefore delayed to time
\begin{equation}
\eta_{\rm rec} \rightarrow \eta_{\rm rec}\, e^{\Phi_L/3},
\end{equation}  which exactly compensates the effect of the time transformation. This amounts to the fact that we effectively only need to rescale $\vec{k}$ to account for a long-wavelength mode. 

We restrict ourselves to the case $n_T=0$ for now and derive the results in two limits:
\begin{itemize}
\item  $10 \lesssim l \lesssim 50$: 
for $l < \eta_{\rm obs}/\eta_{\rm rec}$ one can extend the lower integral bound to zero and obtain 
\begin{equation}
C_l^{BB} \simeq \frac{18 \pi}{25} \left(\frac{\Delta \eta_{\rm rec}}{\eta_{\rm obs}}\right)^2 A_T \int_0^{\infty} \frac{dy}{y^3}\,j_2^2(y),
\end{equation}
where we have made the change of variable $y=k\eta_{\rm rec}$. The integral above equals $1/72$.
From Eq. (\ref{eq:resc}) we deduce that the effect on $C_l^{BB}$ is 
\begin{equation}
C_l^{BB} \rightarrow C_l^{BB} (1 + (5 \times 2) a(l_3) ).
\end{equation}
On the other hand, by rescaling the power spectrum we obtain
\begin{eqnarray}
C_l^{BB} &\simeq & \frac{18 \pi}{25} \Delta \eta_{\rm rec}^2 \int_{(l+1/2)/ \eta_{\rm obs}}^{\infty} d \ln k \,\, A_T \frac{j_2^2(e^{5\Phi_L /3} k \eta_{\rm rec})}{\eta_{\rm rec}^2} \left[ \frac{1}{(k\eta_{\rm obs})^2}\right]\nonumber\\
&=&\frac{18 \pi}{25} \left(\frac{\Delta \eta_{\rm rec}}{\eta_{\rm obs}}\right)^2 A_T \int_0^{\infty} \frac{dy}{y^3}\,j_2^2(y) \times e^{10\Phi_L /3}
\end{eqnarray}
where we defined $y= k\, e^{5\Phi_L /3} /\eta_{\rm rec}$. This produces the same effect at first order as expected.

\begin{figure*}[h!t]
\centering
\includegraphics[width=.65\textwidth]{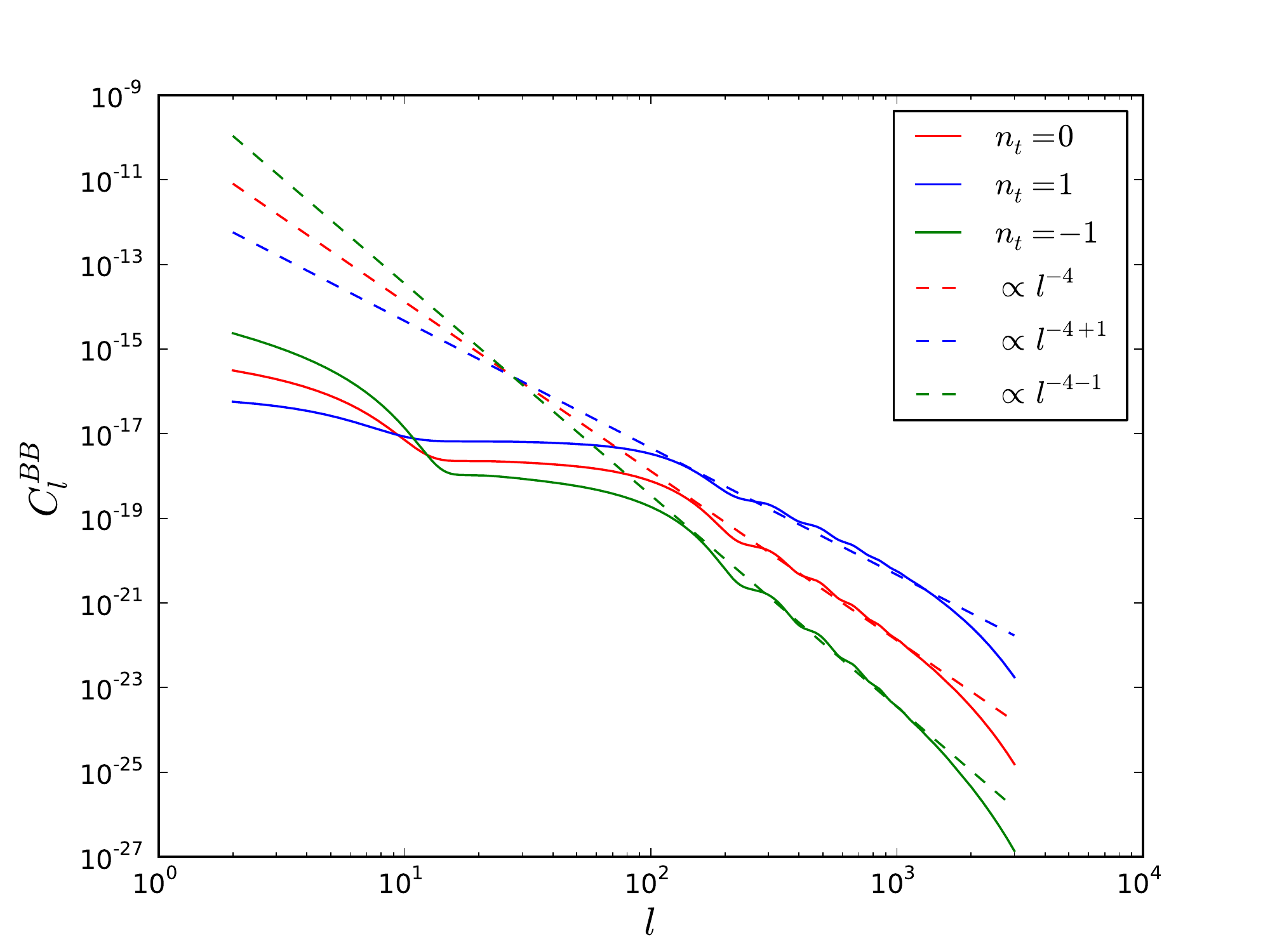}
\caption{Comparison of the slope of the power spectrum with the analytical approximation for the derivative as function of $n_T$.}
\label{fig:derivative}
\end{figure*}

\item $50 \lesssim l \lesssim 10^3$: in this regime, one cannot extend the lower integral bound to zero. The main contribution to the integral comes from modes $k \sim l / \eta_{\rm obs} $ so $k \eta_{\rm rec} \gg 1$ and we can use the large argument limit of the transfer function
\begin{equation}
\frac{j_2^2(k \eta_{\rm rec})}{\eta_{\rm rec}^2} \simeq \frac{\sin^2((k\eta_{\rm rec})^2)}{\eta_{\rm rec}^2(k\eta_{\rm rec})^2}
\end{equation}
leading to 
\begin{equation}
C_l^{BB} \simeq \frac{36 \pi}{25} \left(\frac{\Delta \eta_{\rm rec}}{\eta_{\rm rec}}\right)^2 A_T \int_{(l+1/2)/ \eta_{\rm obs}}^{\infty} d \ln k \,\,  \frac{\sin^2((k\eta_{\rm rec})^2)}{(k\eta_{\rm rec})^2} \left[ \frac{\sqrt{(k\eta_{\rm obs})^2-(l+1/2)^2}}{2(k\eta_{\rm obs})^3}\right].
\end{equation}
The sine can be averaged and upon defining $u= k \eta_{\rm obs} / (l+1/2)$ one obtains
\begin{equation}
C_l^{BB} \simeq \frac{9 \pi}{25} A_T \left(\frac{\Delta \eta_{\rm rec}}{\eta_{\rm rec}}\right)^2 \left(\frac{\eta_{\rm obs}}{\eta_{\rm rec}}\right)^2  \frac{1}{(l+1/2)^4}\int_{1}^{\infty}  \frac{du}{u^6} \,\,  \sqrt{u^2-1}.
\end{equation}
The integral equals $2/15$. From Eq. (\ref{eq:resc}), we deduce that the effect on $C_l^{BB}$ is 
\begin{equation}
C_l^{BB} \rightarrow C_l^{BB} (1 + (5 \times (-2)) a(l_3) ),
\end{equation}
while the rescaling acts only on the transfer function  $j_2 \propto k^{-2}$ which produces 
\begin{equation}
C_l^{BB} \rightarrow C_l^{BB} e^{-10 \Phi_L /3} \simeq  C_l^{BB}(1-10a(l_3)).
\end{equation}
\end{itemize}
Finally, one can compute the effect for $n_T \neq 0$ from the rescaling easily. Following the same procedure we obtain
\bea \label{eq:nt}
C_l^{BB}  \rightarrow  C_l^{BB} & (1 + 10 a(l_3) )   & 10 \lesssim  l  \lesssim 50,\\
C_l^{BB}  \rightarrow C_l^{BB} &(1 + 5 (n_T-2) a(l_3) )    & 50 \lesssim  l \lesssim 10^3,
\eea
which also coincides with the result computed using Eq. \eqref{eq:resc}.
The above expressions can be confirmed by numerical evaluations of the logarithmic derivative $d \ln (l^2 C_l)/d \ln l$. Figure (\ref{fig:derivative}) shows a qualitative agreement between the numerical and analytical behavior in the two regimes, namely $ d \ln (l^2 C_l)/d \ln l =2$ and $ d \ln (l^2 C_l)/d \ln l = ( n_T-2)$ for low and high $l$ respectively.

\section{Planck, COrE and PRISM instrumental characteristics} 
\begin{table}[ht]\label{tab:Planck}
\centering
    \begin{tabular}{ c | c | c | c } 
         Frequency  \ & $\theta_{\rm fwhm}$ \ & \ $\sigma_T$ \ &  \ $\sigma_P \ $ \\ 
         \ (Hz)  \ & \ $({\rm arcmin})$ \ & \ \ $(\mu K/K) \ $ & \ \ $(\mu K/K)$ \\
     \hline
        100 &  10 & 2.5 & 4.0 \\
        143 & 7.1 & 2.2 & 4.2 \\ 
        217 &  5.0 & 4.8 & 9.8  \\
    \end{tabular}
    \caption{Planck \ (14-month mission)}
\end{table} 
\begin{table}[ht]\label{tab:COrE}
\centering
    \begin{tabular}{c|  c |c| c}  
         Frequency  \ & $\theta_{\rm fwhm}$ \ & \ $\sigma_T$ \ &  \ $\sigma_P \ $ \\ 
         \ (Hz)  \ & \ $({\rm arcmin})$ \ & \ \ $(\mu K) \ $ & \ \ $(\mu K)$ \\
        \hline 
        105 &  10 & 0.268 & 0.463 \\
        135 & 7.8 & 0.338 & 0.583 \\ 
        165 & 6.4 & 0.417 &  0.72 \\
        195 & 5.4 & 0.487 & 0.841\\
        225 & 4.7 & 0.562 & 0.972 \\
        255 & 4.1 & 1.48   & 2.56\\
        285 & 3.7 & 2.73 & 4.7 \\
    \end{tabular}
    \caption{COrE \ (4-year mission)}
\end{table}
\begin{table}[ht]\label{tab:PRISM}
\centering
\begin{tabular}{c |c |c |c}
 Frequency  \ & $\theta_{\rm fwhm}$ \ & \ $\sigma_T$ \ &  \ $\sigma_P \ $ \\ 
         \ (Hz)  \ & \ $({\rm arcmin})$ \ & \ \ $(\mu K) \ $ & \ \ $(\mu K)$ \\
\hline 
      105   &  4.8 &  0.601 &  0.849\\   
      135   & 3.8 &  0.682 &  0.963 \\   
      160   &  3.2 &  0.760 &  1.074  \\   
      185   &  2.8 &  0.899 &  1.27  \\   
      200   &  2.5 &  1.03 &  1.47 \\   
\end{tabular}
\caption{ PRISM (4-year mission)}
\end{table}

\newpage

\bibliography{SCINC}

\begingroup\raggedright\begin{thebibliography}{40}
\expandafter\ifx\csname natexlab\endcsname\relax\def\natexlab#1{#1}\fi

\bibitem[Ade et~al.(2014)]{Ade:2014xna}
{\bfseries BICEP2 Collaboration} Collaboration, P.~Ade {\em et~al.},
  ``{Detection of B-Mode Polarization at Degree Angular Scales by BICEP2}'',
  {\em Phys.Rev.Lett.} {\bfseries 112} (2014) 241101,
 \href{http://xxx.lanl.gov/abs/1403.3985}{{\ttfamily arXiv:1403.3985}}.

\bibitem[Flauger et~al.(2014)Flauger, Hill, and Spergel]{Flauger:2014qra}
R.~Flauger, J.~C. Hill, and D.~N. Spergel, ``{Toward an Understanding of
  Foreground Emission in the BICEP2 Region}'',
 \href{http://xxx.lanl.gov/abs/1405.7351}{{\ttfamily arXiv:1405.7351}}.

\bibitem[Mortonson and Seljak(2014)]{Mortonson:2014bja}
M.~J. Mortonson and U.~Seljak, ``{A joint analysis of Planck and BICEP2 B modes
  including dust polarization uncertainty}'',
 \href{http://xxx.lanl.gov/abs/1405.5857}{{\ttfamily arXiv:1405.5857}}.

\bibitem[Maldacena(2003)]{Maldacena:2002vr}
J.~M. Maldacena, ``{Non-Gaussian features of primordial fluctuations in single
  field inflationary models}'', {\em JHEP} {\bfseries 0305} (2003) 013,
 \href{http://xxx.lanl.gov/abs/astro-ph/0210603}{{\ttfamily
  arXiv:astro-ph/0210603}}.

\bibitem[Acquaviva et~al.(2003)Acquaviva, Bartolo, Matarrese, and
  Riotto]{Acquaviva:2002ud}
V.~Acquaviva, N.~Bartolo, S.~Matarrese, and A.~Riotto, ``{Second order
  cosmological perturbations from inflation}'', {\em Nucl.Phys.} {\bfseries
  B667} (2003) 119--148,
 \href{http://xxx.lanl.gov/abs/astro-ph/0209156}{{\ttfamily
  arXiv:astro-ph/0209156}}.

\bibitem[Creminelli and Zaldarriaga(2004)]{Creminelli:2004yq}
P.~Creminelli and M.~Zaldarriaga, ``{Single field consistency relation for the
  3-point function}'', {\em JCAP} {\bfseries 0410} (2004) 006,
 \href{http://xxx.lanl.gov/abs/astro-ph/0407059}{{\ttfamily
  arXiv:astro-ph/0407059}}.

\bibitem[Creminelli et~al.(2011)Creminelli, D'Amico, Musso, and
  Nore\~na]{Creminelli:2011rh}
P.~Creminelli, G.~D'Amico, M.~Musso, and J.~Nore\~na, ``{The (not so) squeezed
  limit of the primordial 3-point function}'', {\em JCAP} {\bfseries 1111}
  (2011) 038,
 \href{http://xxx.lanl.gov/abs/1106.1462}{{\ttfamily arXiv:1106.1462}}.

\bibitem[Creminelli et~al.(2012)Creminelli, Nore\~na, and
  Simonovi\'c]{Creminelli:2012ed}
P.~Creminelli, J.~Nore\~na, and M.~Simonovi\'c, ``{Conformal consistency
  relations for single-field inflation}'', {\em JCAP} {\bfseries 1207} (2012)
  052,
 \href{http://xxx.lanl.gov/abs/1203.4595}{{\ttfamily arXiv:1203.4595}}.

\bibitem[Kehagias and Riotto(2012)]{Kehagias:2012pd}
A.~Kehagias and A.~Riotto, ``{Operator Product Expansion of Inflationary
  Correlators and Conformal Symmetry of de Sitter}'', {\em Nucl.Phys.}
  {\bfseries B864} (2012) 492--529,
 \href{http://xxx.lanl.gov/abs/1205.1523}{{\ttfamily arXiv:1205.1523}}.

\bibitem[Assassi et~al.(2012)Assassi, Baumann, and Green]{Assassi:2012zq}
V.~Assassi, D.~Baumann, and D.~Green, ``{On Soft Limits of Inflationary
  Correlation Functions}'', {\em JCAP} {\bfseries 1211} (2012) 047,
 \href{http://xxx.lanl.gov/abs/1204.4207}{{\ttfamily arXiv:1204.4207}}.

\bibitem[Hinterbichler et~al.(2014)Hinterbichler, Hui, and
  Khoury]{Hinterbichler:2013dpa}
K.~Hinterbichler, L.~Hui, and J.~Khoury, ``{An Infinite Set of Ward Identities
  for Adiabatic Modes in Cosmology}'', {\em JCAP} {\bfseries 1401} (2014) 039,
 \href{http://xxx.lanl.gov/abs/1304.5527}{{\ttfamily arXiv:1304.5527}}.

\bibitem[Creminelli et~al.(2011)Creminelli, Pitrou, and
  Vernizzi]{Creminelli:2011sq}
P.~Creminelli, C.~Pitrou, and F.~Vernizzi, ``{The CMB bispectrum in the
  squeezed limit}'', {\em JCAP} {\bfseries 1111} (2011) 025,
 \href{http://xxx.lanl.gov/abs/1109.1822}{{\ttfamily arXiv:1109.1822}}.

\bibitem[Bartolo et~al.(2012)Bartolo, Matarrese, and Riotto]{Bartolo:2011wb}
N.~Bartolo, S.~Matarrese, and A.~Riotto, ``{Non-Gaussianity in the Cosmic
  Microwave Background Anisotropies at Recombination in the Squeezed limit}'',
  {\em JCAP} {\bfseries 1202} (2012) 017,
 \href{http://xxx.lanl.gov/abs/1109.2043}{{\ttfamily arXiv:1109.2043}}.

\bibitem[Lewis(2012)]{Lewis:2012tc}
A.~Lewis, ``{The full squeezed CMB bispectrum from inflation}'', {\em JCAP}
  {\bfseries 1206} (2012) 023,
 \href{http://xxx.lanl.gov/abs/1204.5018}{{\ttfamily arXiv:1204.5018}}.

\bibitem[Peloso and Pietroni(2013)]{Peloso:2013zw}
M.~Peloso and M.~Pietroni, ``{Galilean invariance and the consistency relation
  for the nonlinear squeezed bispectrum of large scale structure}'', {\em JCAP}
  {\bfseries 1305} (2013) 031,
 \href{http://xxx.lanl.gov/abs/1302.0223}{{\ttfamily arXiv:1302.0223}}.

\bibitem[Kehagias and Riotto(2013)]{Kehagias:2013yd}
A.~Kehagias and A.~Riotto, ``{Symmetries and Consistency Relations in the Large
  Scale Structure of the Universe}'', {\em Nucl.Phys.} {\bfseries B873} (2013)
  514--529,
 \href{http://xxx.lanl.gov/abs/1302.0130}{{\ttfamily arXiv:1302.0130}}.

\bibitem[Creminelli et~al.(2013)Creminelli, Nore\~na, Simonovi\'c, and
  Vernizzi]{Creminelli:2013mca}
P.~Creminelli, J.~Nore\~na, M.~Simonovi\'c, and F.~Vernizzi, ``{Single-Field
  Consistency Relations of Large Scale Structure}'', {\em JCAP} {\bfseries
  1312} (2013) 025,
 \href{http://xxx.lanl.gov/abs/1309.3557}{{\ttfamily arXiv:1309.3557}}.

\bibitem[Kehagias and Riotto(2014)]{Kehagias:2013xga}
A.~Kehagias and A.~Riotto, ``{Conformal Symmetries of FRW Accelerating
  Cosmologies}'', {\em Nucl.Phys.} {\bfseries B884} (2014) 547--565,
 \href{http://xxx.lanl.gov/abs/1309.3671}{{\ttfamily arXiv:1309.3671}}.

\bibitem[Kehagias et~al.(2014)Kehagias, Nore\~na, Perrier, and
  Riotto]{Kehagias:2013rpa}
A.~Kehagias, J.~Nore\~na, H.~Perrier, and A.~Riotto, ``{Consequences of
  Symmetries and Consistency Relations in the Large-Scale Structure of the
  Universe for Non-local bias and Modified Gravity}'', {\em Nucl.Phys.}
  {\bfseries B883} (2014) 83--106,
 \href{http://xxx.lanl.gov/abs/1311.0786}{{\ttfamily arXiv:1311.0786}}.

\bibitem[Valageas(2013)]{Valageas:2013cma}
P.~Valageas, ``{Consistency relations of large-scale structures}'',
 \href{http://xxx.lanl.gov/abs/1311.1236}{{\ttfamily arXiv:1311.1236}}.

\bibitem[Creminelli et~al.(2014)Creminelli, Gleyzes, Simonovi\'c, and
  Vernizzi]{Creminelli:2013poa}
P.~Creminelli, J.~Gleyzes, M.~Simonovi\'c, and F.~Vernizzi, ``{Single-Field
  Consistency Relations of Large Scale Structure. Part II: Resummation and
  Redshift Space}'', {\em JCAP} {\bfseries 1402} (2014) 051,
 \href{http://xxx.lanl.gov/abs/1311.0290}{{\ttfamily arXiv:1311.0290}}.

\bibitem[Valageas(2014)]{Valageas:2013zda}
P.~Valageas, ``{Angular averaged consistency relations of large-scale
  structures}'', {\em Phys.Rev.} {\bfseries D89} (2014) 123522,
 \href{http://xxx.lanl.gov/abs/1311.4286}{{\ttfamily arXiv:1311.4286}}.

\bibitem[Creminelli et~al.(2014)Creminelli, Gleyzes, Hui, Simonovi\'c, and
  Vernizzi]{Creminelli:2013nua}
P.~Creminelli, J.~Gleyzes, L.~Hui, M.~Simonovi\'c, and F.~Vernizzi,
  ``{Single-Field Consistency Relations of Large Scale Structure. Part III:
  Test of the Equivalence Principle}'', {\em JCAP} {\bfseries 1406} (2014) 009,
 \href{http://xxx.lanl.gov/abs/1312.6074}{{\ttfamily arXiv:1312.6074}}.

\bibitem[Nishimichi and Valageas(2014)]{Nishimichi:2014jna}
T.~Nishimichi and P.~Valageas, ``{Testing the equal-time angular-averaged
  consistency relation of the gravitational dynamics in N-body simulations}'',
 \href{http://xxx.lanl.gov/abs/1402.3293}{{\ttfamily arXiv:1402.3293}}.

\bibitem[Horn et~al.(2014)Horn, Hui, and Xiao]{Horn:2014rta}
B.~Horn, L.~Hui, and X.~Xiao, ``{Soft-Pion Theorems for Large Scale
  Structure}'',
 \href{http://xxx.lanl.gov/abs/1406.0842}{{\ttfamily arXiv:1406.0842}}.

\bibitem[Boubekeur et~al.(2009)Boubekeur, Creminelli, D'Amico, Nore\~na, and
  Vernizzi]{Boubekeur:2009uk}
L.~Boubekeur, P.~Creminelli, G.~D'Amico, J.~Nore\~na, and F.~Vernizzi,
  ``{Sachs-Wolfe at second order: the CMB bispectrum on large angular
  scales}'', {\em JCAP} {\bfseries 0908} (2009) 029,
 \href{http://xxx.lanl.gov/abs/0906.0980}{{\ttfamily arXiv:0906.0980}}.

\bibitem[Huang and Vernizzi(2014)]{Huang:2013qua}
Z.~Huang and F.~Vernizzi, ``{The full CMB temperature bispectrum from
  single-field inflation}'', {\em Phys.Rev.} {\bfseries D89} (2014) 021302,
 \href{http://xxx.lanl.gov/abs/1311.6105}{{\ttfamily arXiv:1311.6105}}.

\bibitem[Pettinari et~al.(2014)Pettinari, Fidler, Crittenden, Koyama, Lewis,
  et~al.]{Pettinari:2014iha}
G.~W. Pettinari, C.~Fidler, R.~Crittenden, K.~Koyama, A.~Lewis, {\em et~al.},
  ``{Impact of polarisation on the intrinsic CMB bispectrum}'',
 \href{http://xxx.lanl.gov/abs/1406.2981}{{\ttfamily arXiv:1406.2981}}.

\bibitem[Weinberg(2003)]{Weinberg:2003sw}
S.~Weinberg, ``{Adiabatic modes in cosmology}'', {\em Phys.Rev.} {\bfseries
  D67} (2003) 123504,
 \href{http://xxx.lanl.gov/abs/astro-ph/0302326}{{\ttfamily
  arXiv:astro-ph/0302326}}.

\bibitem[Bartolo et~al.(2006)Bartolo, Matarrese, and Riotto]{Bartolo:2006cu}
N.~Bartolo, S.~Matarrese, and A.~Riotto, ``{CMB Anisotropies at Second Order
  I}'', {\em JCAP} {\bfseries 0606} (2006) 024,
 \href{http://xxx.lanl.gov/abs/astro-ph/0604416}{{\ttfamily
  arXiv:astro-ph/0604416}}.

\bibitem[Bartolo et~al.(2007)Bartolo, Matarrese, and Riotto]{Bartolo:2006fj}
N.~Bartolo, S.~Matarrese, and A.~Riotto, ``{CMB Anisotropies at Second-Order.
  2. Analytical Approach}'', {\em JCAP} {\bfseries 0701} (2007) 019,
 \href{http://xxx.lanl.gov/abs/astro-ph/0610110}{{\ttfamily
  arXiv:astro-ph/0610110}}.

\bibitem[Zaldarriaga and Seljak(1997)]{Zaldarriaga:1996xe}
M.~Zaldarriaga and U.~Seljak, ``{An all sky analysis of polarization in the
  microwave background}'', {\em Phys.Rev.} {\bfseries D55} (1997) 1830--1840,
 \href{http://xxx.lanl.gov/abs/astro-ph/9609170}{{\ttfamily
  arXiv:astro-ph/9609170}}.

\bibitem[Hu(2000)]{Hu:2000ee}
W.~Hu, ``{Weak lensing of the CMB: A harmonic approach}'', {\em Phys.Rev.}
  {\bfseries D62} (2000) 043007,
 \href{http://xxx.lanl.gov/abs/astro-ph/0001303}{{\ttfamily
  arXiv:astro-ph/0001303}}.

\bibitem[Bowden et~al.(2004)Bowden, Taylor, Ganga, Ade, Bock,
  et~al.]{Bowden:2003ub}
M.~Bowden, A.~Taylor, K.~Ganga, P.~Ade, J.~Bock, {\em et~al.}, ``{Scientific
  optimization of a ground - based CMB polarization experiment}'', {\em
  Mon.Not.Roy.Astron.Soc.} {\bfseries 349} (2004) 321,
 \href{http://xxx.lanl.gov/abs/astro-ph/0309610}{{\ttfamily
  arXiv:astro-ph/0309610}}.

\bibitem[Tauber et~al.(2006)]{Planck:2006aa}
{\bfseries Planck Collaboration} Collaboration, J.~Tauber {\em et~al.}, ``{The
  Scientific programme of Planck}'',
 \href{http://xxx.lanl.gov/abs/astro-ph/0604069}{{\ttfamily
  arXiv:astro-ph/0604069}}.

\bibitem[Bouchet et~al.(2011)]{Bouchet:2011ck}
{\bfseries COrE Collaboration} Collaboration, F.~Bouchet {\em et~al.}, ``{COrE
  (Cosmic Origins Explorer) A White Paper}'',
 \href{http://xxx.lanl.gov/abs/1102.2181}{{\ttfamily arXiv:1102.2181}}.

\bibitem[Andre et~al.(2013)]{Andre:2013afa}
{\bfseries PRISM Collaboration} Collaboration, P.~Andre {\em et~al.}, ``{PRISM
  (Polarized Radiation Imaging and Spectroscopy Mission): A White Paper on the
  Ultimate Polarimetric Spectro-Imaging of the Microwave and Far-Infrared
  Sky}'',
 \href{http://xxx.lanl.gov/abs/1306.2259}{{\ttfamily arXiv:1306.2259}}.

\bibitem[Lesgourgues(2011)]{Lesgourgues:2011re}
J.~Lesgourgues, ``{The Cosmic Linear Anisotropy Solving System (CLASS) I:
  Overview}'',
 \href{http://xxx.lanl.gov/abs/1104.2932}{{\ttfamily arXiv:1104.2932}}.

\bibitem[Chiang et~al.(2014)Chiang, Wagner, Schmidt, and
  Komatsu]{Chiang:2014oga}
C.-T. Chiang, C.~Wagner, F.~Schmidt, and E.~Komatsu, ``{Position-dependent
  power spectrum of the large-scale structure: a novel method to measure the
  squeezed-limit bispectrum}'', {\em JCAP} {\bfseries 1405} (2014) 048,
 \href{http://xxx.lanl.gov/abs/1403.3411}{{\ttfamily arXiv:1403.3411}}.

\bibitem[Gorbunov and Rubakov(2011)]{Gorbunov:2011zzc}
D.~S. Gorbunov and V.~A. Rubakov, ``{Introduction to the theory of the early
  universe: Cosmological perturbations and inflationary theory}'',
2011.

\end{thebibliography}\endgroup

\end{document}